\documentclass[twocolumn,preprintnumbers,amsmath,amssymb]{revtex4}
\usepackage{docs}

\usepackage{graphicx}
\usepackage{dcolumn}

\usepackage{bm}

\begin{document}
\title{Nature of time and causality in Physics}%

\author{Francisco S. N. Lobo}%
\email{flobo@cosmo.fis.fc.ul.pt} \affiliation{Centro de Astronomia
e Astrof\'{\i}sica da Universidade de Lisboa, Campo Grande, Ed. C8
1749-016 Lisboa, Portugal}
%
\affiliation{Institute of
Gravitation \& Cosmology, University of Portsmouth, Portsmouth PO1
2EG, UK}
%

\begin{abstract}

The conceptual definition and understanding of the nature of time,
both qualitatively and quantitatively is of the utmost difficulty
and importance, and plays a fundamental role in physics. Physical
systems seem to evolve in paths of increasing entropy and of
complexity, and thus, the arrow of time shall be explored in the
context of thermodynamic irreversibility and quantum physics. In
Newtonian physics, time flows at a constant rate, the same for all
observers; however, it necessarily flows at different rates for
different observers in special and general relativity. Special
relativity provides important quantitative elucidations of the
fundamental processes related to time dilation effects, and
general relativity provides a deep analysis of effects of time
flow, such as in the presence of gravitational fields. Through the
special theory of relativity, time became intimately related with
space, giving rise to the notion of spacetime, in which both
parameters cannot be considered as separate entities. As time is
incorporated into the proper structure of the fabric of spacetime,
it is interesting to note that general relativity is contaminated
with non-trivial geometries that generate closed timelike curves,
and thus apparently violates causality. The notion of causality is
fundamental in the construction of physical theories; therefore
time travel and its associated paradoxes have to be treated with
great caution.
These issues are briefly analyzed in this review paper.

\end{abstract}


\maketitle

\section{Introduction}

Time is a mysterious ingredient of the Universe and stubbornly
resists simple definition. St. Augustine, in his Confessions,
reflecting on the nature of time, states: ``What then is time? If
no one asks me, I know: if I wish to explain it to one that
asketh, I know not'' (The Confessions, ch. XI, sec. 17). Perhaps
the reason for being so illusive is that being a fundamental
quantity, there is nothing more fundamental to be defined in terms
of. Citing Alfred North Whitehead, the philosopher-mathematician:
``It is impossible to mediate on time ... without an overwhelming
emotion at the limitations of human intelligence.'' However,
intuitively, we do verify that the notion of time emerges through
an intimate relationship to change, and subjectively may be
considered as something that flows. This view can be traced back
as far as Aristotle, a keen natural philosopher, who stated that
``time is the measure of change.'' Throughout history, one may
find a wide variety of reflections and considerations on time,
dating back to ancient religions. For instance, a linear notion of
time may be encountered in the Hebrew and the Zoroastrian Iranian
writings, and in the Judaeo-Christian doctrine, based on the Bible
and the unique character of historical events, which possesses a
beginning, namely, the act of creation. In ancient Greece the
image of Chronos, the Father Time, was conveyed, and Plato further
assumed the notion of a circular time, where the latter had a
beginning and looped back unto itself. This was probably inspired
in the cyclic phenomena observed in Nature, namely the alternation
of day and night, the repetition of the seasons, etc. Eastern
religions also have a cyclic notion of time, consisting of a
repetition of births and extinctions. However, it was only in the
17th century that the philosopher Francis Bacon clearly formulated
the concept of linear time, and through the influence of Newton,
Barrow, Leibniz, Locke and Kant amongst others, by the 19th
century the idea of linear time dominated both in science and
philosophy.

In a scientific context, it is perhaps fair to state that
reflections on time culminated in Newton's concept of absolute
time, which assumed that time flowed at the same rate for all
observers in the Universe. Quoting Newton from his {\it Principia}
\cite{Newton}: ``Absolute, true, and mathematical time, in and of
itself and of its own nature, without reference to anything
external, flows uniformly and by another name is called
duration.'' However, in 1905, Albert Einstein changed altogether
our notion of time, through the formulation of the special theory
of relativity and stating, in particular, that time flowed at
different rates for different observers. Three years later,
Hermann Minkowski formally united the parameters of time and
space, giving rise to the notion of a fundamental four-dimensional
entity, spacetime. Citing Minkowski: ``Henceforth space by itself,
and time by itself, are doomed to fade away into mere shadows, and
only a kind of union of the two will preserve an independent
reality.''

If we consider that time is empirically related to change, which
is a variation or sequence of occurrences, then, intuitively, the
latter provides us with a notion of something that flows, and thus
the emergent character of time. In Relativity, the above empirical
notion of a sequence of occurrences is substituted by a sequence
of {\it events}. The concept of an event is an idealization of a
point in space and an instant in time. It is interesting to note
that the concept of an instant, associated to that of {\it
duration}, or an interval of time, is also an extremely subtle
issue, and deserves a brief analysis. One may consider a duration
as an ordered set of instants, contrary to being a sum of
instants. A duration is infinitely divisible into more durations,
and not into an instant. Now the classical concept of time being a
linear continuum implies that between any infinitesimally
neighboring instants, an infinity of instants exist, and the flow
of instants constituting a linear continuum of time is reminiscent
of Zeno's paradoxes. Perhaps the problem can be surpassed by
quantizing time in units of the Planck time, $10^{-43}\,{\rm s}$,
in an eventual theory of quantum gravity. Now, a sequence of
events has a determined temporal order, which is experimentally
verified as specific events -- effects -- are triggered off by
others -- causes -- thus providing us with the notion of
causality.

In the literature, one may find two exclusively mutual concepts of
time, which can be characterized as the relational theories and
the absolute theories of time. The latter imply that time exists
independently of physical spacetime events, contrary to relational
theories that defend that time is but a mere relationship of the
causal ordering of events, i.e., time is an abstract concept,
non-existent as a physical entity, but useful in describing
processes. In particular, an example of an absolute theory of time
can be traced back to early 17th century, with Isaac Barrow's
refutal of the Aristotelian notion that time is related to change,
stating that time is an entity which exists independently of
change or motion. This is reflected in his {\it Lectiones
Geometricae}, in 1676, where he states: ``Whether things run or
stand still, whether we sleep or wake, time flows in its even
tenor.'' His student, Isaac Newton, extended this idea and
compared time and space with an infinitely large vessel,
containing events, and existing independently of the latter. This
notion was, in turn, refuted by Gottfried Leibniz, who defended a
relationship between time and an ordering of non-simultaneous
events: ``Time is the order of possibilities which cannot coexist
and therefore must exist successfully.''

Now, for an emergent subjective notion of time to occur, it seems
that a changing configuration of matter is necessary. For
instance, in an empty universe, a hypothetical observer cannot
measure time nor length, i.e., in a universe without processes one
may argue that the observer cannot experience an emergent notion
of time, or for that matter, of space. However, the absolute time
theorists defend that the container spacetime, i.e., space and
time, still exists. The fundamental question is whether time does
exist independently, in the absence of change. Albert Einstein
seems to provide the controversial answer: ``Till now it was
believed that time and space existed themselves, even if there was
nothing -- no Sun, no Earth, no stars -- while now we know that
time and space are not the vessels for the Universe, but could not
exist at all if there were no contents, namely, no Sun, no Earth,
and other celestial bodies'' (New York Times, 3 December 1919).
But, in another stage of his life contradicts himself, by stating:
``The conceptions of time and space have been such that if
everything in the Universe were taken away, if there were nothing
left, there would still be left to man time and space'' (New York
Times, 4 April 1921). Note that the above ideas further complicate
the issue of the flow of time. Some theorists deny the objective
flow of time, but nevertheless admit the presence of change, and
argue that the empirical notion of the flow is merely subjective.
Their opponents defend an objective flow of time, where events
change from being indeterminate in the future to being determinate
in the present and past.

An interesting example of an absolute theory of time is the
``Block Universe'' description of spacetime as an unchanging
four-dimensional block, where time is considered a dimension. In
this representation, a preferred `now' is non-existent and past
and future times are equally present. All points in time are
equally valid frames of reference, and whether a specific instant
is in the future or past is frame dependent. However, despite the
fact that each observer does indeed experience a subjective flow
of time, special relativity denies the possibility of universal
simultaneity (which shall be treated in more detail below), and
thus the possibility of a universal now. Now, if future events
already exist, why don't we remember the future? We do remember
the past, and this time asymmetry gives rise to a subjective arrow
of time. It appears to the ``Block Universe'' representation that
the notion of the flow of time is a subjective illusion. Note that
the Block Universe point of view inflicts a great blow to the
notion of ``free will'', as it proposes that both past and future
events are as immutably fixed, and consequently impossible to
change. A Block Universe advocate may argue that free will is but
mere determinism in disguise. We refer the reader to Ref.
\cite{Ellis:2006sq} for more details on the objections to the
Block Universe viewpoint.

An important aspect of the nature of time is its arrow. The modern
perspective in physics is that essentially ``dynamical laws''
govern the Universe, namely, given initial conditions of the
physical state, the laws specify the evolution of a determined
physical system with time. However, the dynamical equations of
classical and quantum physics are symmetrical under a time
reversal, i.e., mathematically, one might as well specify the
final conditions and evolve the physical system back in time. But,
several issues are raised by thermodynamics, general relativity
and quantum mechanics on the theme of time asymmetry. In
principle, the latter would enable an observer to empirically
distinguish past from future. For instance, the Second Law of
Thermodynamics, which states that in an isolated system the
entropy (which is a measure of disorder) increases provides a
thermodynamic arrow of time. One may assume that the Second Law of
Thermodynamics and the thermodynamic arrow of time are a
consequence of the initial conditions of the universe, which leads
us to the cosmological arrow of time, that inexorably points in
the direction of the universe's expansion. In the context of
quantum mechanics, a fundamental aspect of the theory is that of
quantum uncertainty, i.e., it is not possible to determine a
unique outcome of quantum events. It is interesting to note that
despite the fact that there is time-symmetry in the evolution of a
quantum system, the reduction of the wave function is essentially
time-asymmetric. These aspects are explored in more detail below.

As time is incorporated into the proper structure of the fabric of
spacetime, it is interesting to note that general relativity is
contaminated with non-trivial geometries that generate closed
timelike curves, and apparently violates causality. A closed
timelike curve allows time travel, in the sense that an observer
who travels on a trajectory in spacetime along this curve, returns
to an event that coincides with the departure. The arrow of time
leads forward, as measured locally by the observer, but globally
he/she may return to an event in the past. This fact apparently
violates causality, opening Pandora's box and producing time
travel paradoxes \cite{Nahin}, throwing a further veil over our
understanding of the fundamental nature of time. The notion of
causality is fundamental in the construction of physical theories;
therefore time travel and its associated paradoxes have to be
treated with great caution \cite{Visser}.

As this chapter is aimed for students and researchers in
Psychology or Neuroscience, the mathematics is kept at a minimum.
We refer the reader to the remaining chapters for the
psychological aspects of time, and only the objective nature of
time will be considered in this chapter, which is outlined in the
following manner: In Section II, the relativistic aspects of time
in special and general relativity will be considered in detail,
where much emphasis will be attributed to spacetime diagrams. In
Section III, time irreversibility and the arrow of time will be
treated in the context of thermodynamics and quantum mechanics. In
Section IV, closed timelike curves and causality violation will be
analyzed, and in Section V, we conclude.

\section{Relativistic time}

The conceptual definition and understanding of time, both
quantitatively and qualitatively is of the utmost difficulty and
importance. Special relativity provides us with important
quantitative elucidations of the fundamental processes related to
time dilation effects. The general theory of relativity provides a
deep analysis to effects of time flow in the presence of strong
and weak gravitational fields. The general theory of relativity
has been an extremely successful theory, with a well established
experimental footing, at least for weak gravitational fields. Its
predictions range from the existence of black holes, gravitational
radiation to the cosmological models predicting a primordial
beginning, namely the big-bang \cite{Hawking,Wald}.

\subsection{Time in special relativity}

To set the stage, perhaps it is important to emphasize that one of
the greatest theoretical triumphs of the 19th century physics was
James Clerk Maxwell's formulation of electromagnetism, which, in
particular, predicted that light waves are electromagnetic in
nature. Now, as was believed, in Maxwell's time, all wave
phenomena required a medium to propagate, and the latter for light
waves was denoted as the ``luminiferous ether.'' Thus, it was
predicted that experiments would allow the absolute motion through
the ether to be detected. However, the famous Michelson-Morley
experiment, devised to measure the velocity of the Earth relative
to the ether came up with a null result. To explain the latter,
Lorentz deduced specific relationships, denoted as the Lorentz
transformations, which are shown below. Einstein also later
derived these transformations in formulating his special theory of
relativity. Explicitly, using the Lorentz transformation, Lorentz
and Fitzgerald explained the Michelson-Morley null result, by
inferring the contraction of rigid bodies and the slowing down of
clocks when moving through the ether. It is also worth mentioning
that the Maxwell equations were not invariant under the Galilean
transformations, i.e., they appeared to violate the Principle of
Galilean Relativity, which essentially states that the dynamical
laws of physics are the same when referred to any uniformally
moving frame. At first, it was thought that Maxwell's equations
were incorrect, and were consequently modified to be invariant
under Galilean transformations. But, this seemed to predict new
electromagnetic phenomena, which could not be experimentally
verified. However, applying the Lorentz transformations, it was
found that the Maxwell equations remain invariant.

\bigskip

To make the above statements more precise, we shall briefly
consider the Galilean transformations, and analyze the Lorentz
transformations in more detail.
Note that the first law of Newtonian physics essentially states
that: ``A body continues in its state of rest or of uniform
motion, unless acted upon by an external force.'' The frame of
reference of a body at rest or in uniform motion, i.e., possessing
a constant velocity, is denoted an {\it inertial} frame.

Consider now an inertial reference frame ${\cal O}'$, with
coordinates $(t',x',y',z')$, moving along the $x$ direction with
uniform velocity, $v$, with respect to another inertial frame
${\cal O}$, with coordinates $(t,x,y,z)$. The Galilean
transformation relates an event in an inertial frame ${\cal O}$ to
another ${\cal O}'$, and are given by the following relationships
\begin{eqnarray}\label{Galilean}
x'&=&x-vt \,,\nonumber \\
y'&=&y\,,\nonumber \\
z'&=&z\,,\nonumber \\
t'&=&t \,.\nonumber
\end{eqnarray}
Note that the last equation is the mathematical assumption of
absolute time in Newtonian physics.

In Euclidean space the distance between two arbitrary points, $A$
and $B$, with coordinates $(t_A,x_A,y_A,z_A)$ and
$(t_B,x_B,y_B,z_B)$, respectively, is given by
\begin{equation}
(\Delta l)^2=(\Delta x)^2+(\Delta y)^2+(\Delta z)^2 \,,
    \label{Euclid}
\end{equation}
where $\Delta x$, $\Delta y$ and $\Delta z$ are the Cartesian
coordinate intervals between $A$ and $B$. One may infer some
interesting properties from the above relationship. First, one
verifies that $\Delta l=0$ if and only if $\Delta x=\Delta
y=\Delta z=0$, which states that both points $A$ and $B$ coincide
when the Euclidean distance between them is zero. Now, one may
show that both the time difference $\Delta t=t_B-t_A$ and the
relationship (\ref{Euclid}) are separately invariant under any
Galilean transformation, which leads one to consider that time and
space are separate entities in Newtonian physics.

\bigskip

However, in 1905, Einstein abandoned the postulate of absolute
time, and assumed the following two postulates: $(i)$ the speed of
light, $c$, is the same in all inertial frames; $(ii)$ the
principle of relativity, which states that the laws of physics
take the same form in every inertial frame. Considering, once
again, an inertial reference frame ${\cal O}'$, with coordinates
$(t',x',y',z')$, moving along the $x$ direction with uniform
velocity relative to another inertial frame ${\cal O}$, with
coordinates $(t,x,y,z)$, and taking into account the above two
postulates, Einstein deduced the Lorentz transformation, which are
given by
\begin{eqnarray}\label{Lorentz}
t'&=&\gamma (t-vx/c^2) \,,\nonumber  \\
x'&=&\gamma (x-vt) \,,\nonumber \\
y'&=&y\,,\nonumber \\
z'&=&z\,,\nonumber
\end{eqnarray}
where $\gamma$ is defined as
\begin{equation}\label{gamma}
\gamma=(1-v^2/c^2)^{-1/2} \,.
\end{equation}
One immediately verifies, from the first two equations, that the
time and space coordinates are mixed by the Lorentz
transformation, and hence, the viewpoint that the physical world
is modelled by a four-dimensional spacetime continuum.

\bigskip

Considering two events, $A$ and $B$, respectively, with
coordinates $(t_A,x_A,y_A,z_A)$ and $(t_B,x_B,y_B,z_B)$ in an
inertial frame ${\cal O}$, then the interval between the events is
given by
\begin{equation}
\Delta s^2=-c^2\Delta t^2+\Delta x^2+\Delta y^2+\Delta z^2 \,,
    \label{SpecRel}
\end{equation}
where $c$ is the speed of light, and $\Delta t$ is the time
interval between the two events $A$ and $B$ \cite{Hobson}. Note
that for this case if $\Delta s=0$, one cannot conclude that
$\Delta t=\Delta x=\Delta y=\Delta z=0$, due to the minus sign
associated with the temporal interval. One verifies that the
expression (\ref{SpecRel}) is invariant under a Lorentz
transformation, and as advocated by Minkowski, space and time are
united in a four-dimensional entity, denoted as spacetime. Thus,
the interval (\ref{SpecRel}) may be considered as an underlying
geometrical property of the spacetime itself. The sign $\Delta
s^2$ is also invariantly defined, so that
\begin{eqnarray}
\Delta s^2&<&0, \qquad {\rm timelike \;\;interval}\,,
   \nonumber   \\
\Delta s^2&=&0, \qquad {\rm null \;\;interval}\,,
  \nonumber   \\
\Delta s^2&>&0, \qquad {\rm spacelike \;\; interval} \,.
   \nonumber
\end{eqnarray}

\begin{figure}[h]
\centering
\includegraphics[width=3.0in]{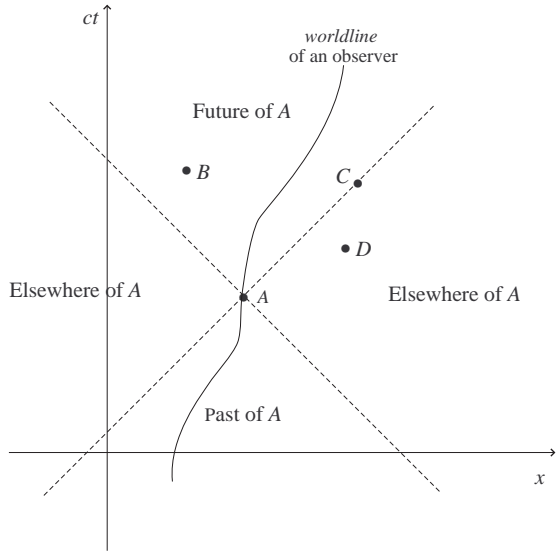}
\caption{Spacetime diagram representing the nature of space and
time, where, for simplicity, the spatial dimensions $y$ and $z$
have been suppressed. Events $A$ and $B$ are separated by a
timelike interval; events $A$ and $C$ by a null interval; and
events $A$ and $D$ are separated by a spacelike interval. The
curve represents the wordline of a timelike observer. The dashed
lines constitute the light cones of $A$. The exterior of the light
cones, denoted the ``elsewhere region of $A$'', constitutes the
region of events which cannot influence nor be influenced by $A$.
See the text for details. } \label{plot1}
\end{figure}

It is useful to represent the nature of space and time using
spacetime diagrams, as depicted in Fig. \ref{plot1}. The diagrams
present a view of the entire spacetime, without a special status
associated to the present time, as will be shown below. For
simplicity, in the spacetime diagrams presented the $y$ and $z$
spatial dimensions have been suppressed. Observers moving with a
relative velocity $v<c$ travel along timelike curves, for instance
as depicted by the curve in Fig. \ref{plot1}, which is denoted by
the {\it worldline} of the observer. Now, there is a unique time
measured along a worldline, denoted as {\it proper time}. A photon
travels along null curves, $ct = \pm x$, which are depicted by the
dashed curves in Fig. \ref{plot1}, and constitute the light cone
of $A$. Events $A$ and $B$ are separated by a timelike interval;
events $A$ and $C$ by a null interval; and events $A$ and $D$ are
separated by a spacelike interval. All events within the upper
light cone of $A$ are in the future of $A$, and all events with
the lower light cone constitute the past of $A$. Events outside
the light cone, such as event $D$ only become visible to $A$ when
it enters the light cone of $A$.

The interior of the future light cone of $A$ constitutes the
region that can be influenced by $A$ with objects travelling with
less than the speed of light. The boundary of the future light
cone can only be influenced by signals with the speed of light
from $A$. In counterpart, the past light cone constitutes the
region in spacetime with events that may influence $A$. The
exterior of the light cones, denoted the ``elsewhere region of
$A$'', constitutes the region of events which cannot influence nor
be influenced by $A$ \cite{Hobson}.

\begin{figure}[h]
\centering
\includegraphics[width=2.4in]{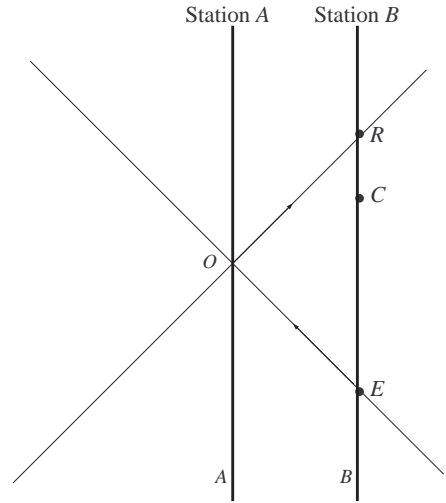}
\caption{Spacetime region representing an example of the
``elsewhere region''. Consider two stationary space-stations $A$
and $B$, where the former observes event $E$, representing a
threatening asteroid on collision course with the station $B$.
However, as the event $C$, depicting the collision of $B$ with the
asteroid is situated outside the causal future of $A$, i.e., in
the ``elsewhere'' region, it is impossible for $A$ to warn $B$ of
the impeding danger in time to avoid the collision. See the text
for details.} \label{elsewhere}
\end{figure}

To illustrate the latter feature, consider the following example
\cite{Ellis}, which is depicted in Fig. \ref{elsewhere}. Assume
the existence of two stationary space-stations, $A$ and $B$,
respectively. $A$ is observing $B$ using a powerful telescope, and
at event $O$ observes event $E$, a threatening asteroid on
collision course with $B$. Despite of sending a warning signal
which arrives at $B$ at event $R$, it will be impossible to warn
$B$ of the impeding danger in time to avoid the collision. This is
due to event $C$, collision $B$-asteroid, being outside the causal
future of $A$, i.e., in the ``elsewhere'' region of $A$. This is
depicted in Fig. \ref{elsewhere}.

\begin{figure}[h]
\centering
\includegraphics[width=3.0in]{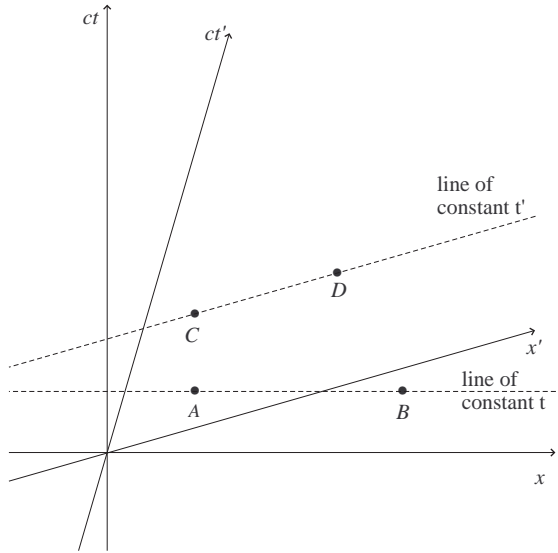}
\caption{Consider an inertial observer ${\cal O}'$, moving along
the $x-$direction with uniform velocity, $v$, relative to another
inertial observer ${\cal O}$. The dashed line parallel to the
$x-$axis represent events for a constant time $t$, such that
events $A$ and $B$ are simultaneous in the observer's ${\cal O}$
reference frame. The dashed line parallel to the $x'-$axis depicts
simultaneous events in observer's ${\cal O}'$ reference frame.
Note that $C$ and $D$ are simultaneous in ${\cal O}'$, but $C$
precedes $D$ for observer ${\cal O}$. Thus, special relativity
denies the possibility of universal simultaneity, and consequently
the possibility of a universal now.} \label{plot2}
\end{figure}

The special theory of relativity challenges many of our intuitive
beliefs about time. For instance, the theory is inconsistent with
the common belief that the temporal order in which two events
occur is independent of the observer's reference frame. To
illustrate this fact, consider an inertial frame ${\cal O}'$, with
coordinates $(t',x')$, moving along the $x$ direction with uniform
velocity, $v$, with respect to another inertial frame ${\cal O}$,
with coordinates $(t,x)$. In Fig. \ref{plot2}, the dashed line
parallel to the $x$-axis represents events for a constant time
$t$, so that events $A$ and $B$ are simultaneous in the observer's
${\cal O}$ reference frame. The dashed line parallel to the
$x'$-axis depicts simultaneous events in observer's ${\cal O}'$
reference frame so that despite the fact that events $C$ and $D$
are simultaneous in ${\cal O}'$, one verifies that $C$ precedes
$D$ for the observer ${\cal O}$.
\bigskip

\begin{figure}[h]
\centering
\includegraphics[width=3.0in]{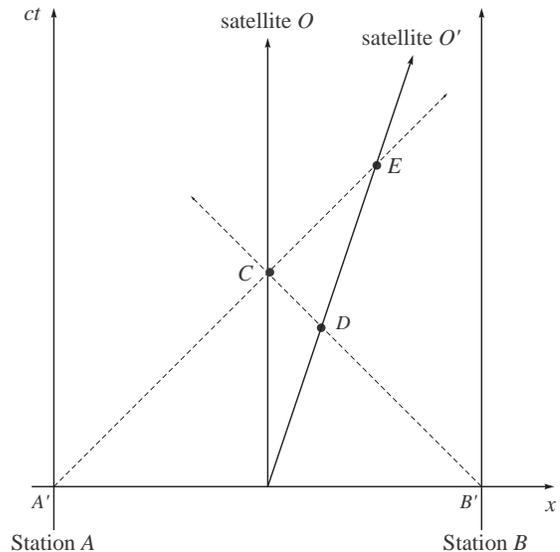}
\caption{Spacetime diagram representing an example of the
impossibility of universal simultaneity. Consider two stationary
space-stations, $A$ and $B$, respectively, a stationary satellite
${\cal O}$ midway between the stations, and another satellite
${\cal O}'$ moving towards $B$ with a velocity $v$ relatively to
$A$. Satellite ${\cal O}$ measures at event $C$ simultaneous
signals, $A'$ and $B'$, sent out by $A$ and $B$, respectively. On
the other hand, satellite ${\cal O}'$ will receive the signal from
$B$, at event $D$, before the signal from $A$, at event $E$.}
\label{simultaneity}
\end{figure}

Consider the following example to further illustrate this point,
which is depicted in Fig. \ref{simultaneity}. Consider two
space-stations $A$ and $B$, separated by a distance $D$. Assume
now a stationary satellite ${\cal O}$ midway between the stations,
and a second satellite ${\cal O}'$ moving towards station $B$ with
a velocity $v$ with respect to ${\cal O}$. At the instant that
both satellites are midway between the stations, these send out a
simultaneous signal, $A'$ and $B'$, respectively, as measured by
$A$ at event $C$. However, satellite ${\cal O}'$ will receive the
signal from station $B$, at event $D$, before the signal from $A$,
event $E$, as depicted in Fig. \ref{simultaneity}.
Thus, whether a specific instant is in the future or past is frame
dependent. {\it Special relativity denies the possibility of
universal simultaneity, and hence the possibility of a universal
now}.
\bigskip

\begin{figure}[h]
\centering
\includegraphics[width=2.0in]{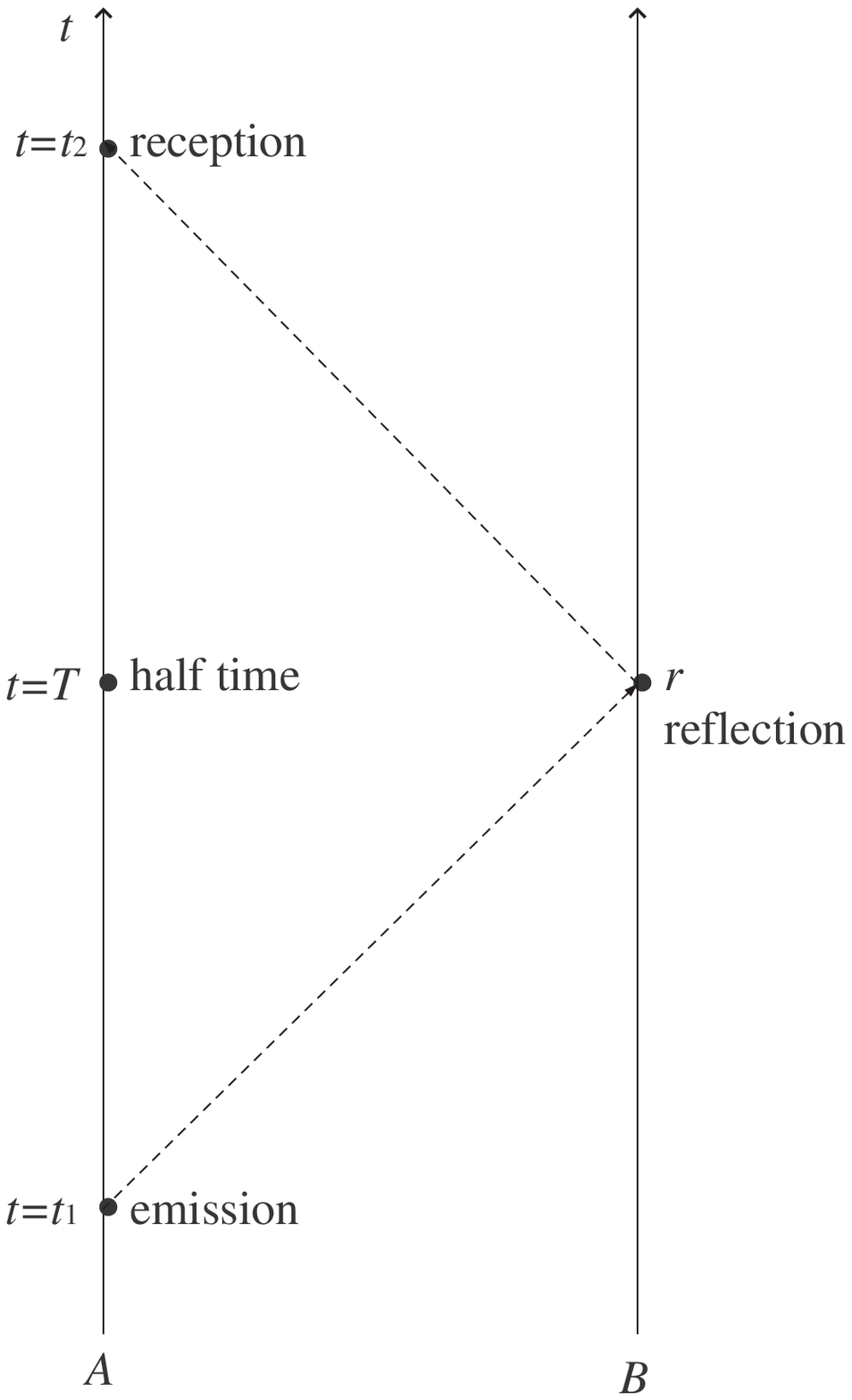}
\caption{Figure depicting the synchronization of distinct clocks
and the definition of simultaneity in spacetime. See the text for
details.} \label{timeclock}
\end{figure}

This raises the problem of the synchronization of distant clocks
in defining simultaneity in spacetime. A simple conceptual
definition of a clock will be provided below. Consider two clocks
at rest with respect with an observer, located at spacetime points
$A$ and $B$, respectively. Suppose now that at time $t_1$, $A$
sends out a signal to $B$, which is reflected and returns to $A$
at time $t_2$. Taking into account the constancy of the speed of
light, $A$ will conclude that the event reflection from $B$, is
simultaneous with the time $T$ is his worldline, which is
precisely half the interval of travel time, i.e.,
\begin{equation}
T=\frac{1}{2}(t_1+t_2)\,.
\end{equation}
This is a simple and practical way of determining simultaneity and
of synchronizing clocks, which is depicted in Fig.
\ref{timeclock}.

\bigskip

Another feature of our intuitive beliefs challenged by the special
theory of relativity is that related with time dilation effects.
For instance, consider the following thought experiment suggested
by Einstein. Suppose that an observer travels along a tram moving
with a relativistic velocity, whilst observing a large clock
through a powerful telescope. The observer sees the clock, as the
emitted light catches up with the tram. Now, if the tram moves at
a speed close to that of light, the light rays emitted from the
clock take longer to catch up with the observer. It seem that time
has slowed down as measured by the latter. If the tram now attains
the speed of light, then the light reflected from the clock cannot
catch up with the observer, and it seems that time would come to a
standstill.

\begin{figure}[h]
\centering
\includegraphics[width=3.4in]{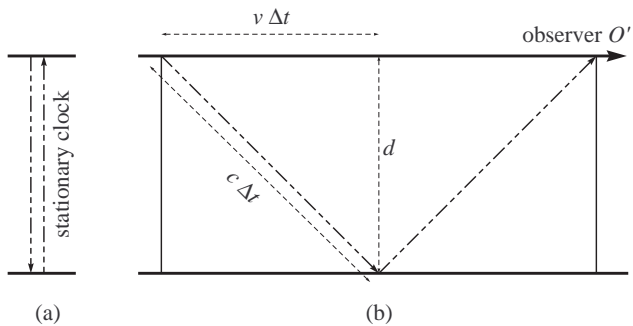}
\caption{$(a)$ A light clock at rest in an inertial reference
frame ${\cal O}'$. The light is reflected from a mirror at a
distance $d$, and is received after a time $2\Delta t'$ as
measured in ${\cal O}'$. $(b)$ This depicts a light clock at rest
with respect to an inertial frame ${\cal O}'$, which is travelling
at a speed $v$ with respect to an identical clock on the ground,
at rest in an inertial frame ${\cal O}$. The latter observer
measures a `click' in a time interval $2\Delta t$. See the text
for details.} \label{clock}
\end{figure}

One may infer time dilation from the constancy of the speed of
light. First it is useful to provide a simple conceptual
definition of a clock, namely, that of a `light clock'. The latter
is constructed by two mirrors separated by a distance $d$, with a
photon being continuously reflected in between. A `click' of this
idealized clock is constituted by the time interval, $2\Delta t'$,
with which the photon traces the distance $2d$, as depicted in
Fig. \ref{clock}. Thus, one deduces the following expression
$2\Delta t'=2d/c$, which implies $\Delta t'=d/c$. Now, consider
that this clock is at rest in an inertial frame ${\cal O}'$
travelling with a relativistic velocity $v$ with respect to a
frame ${\cal O}$. From the special relativistic postulate that the
speed of light is constant in all frames, the time interval traced
out by the photon, as measured by the observer at rest ${\cal O}$
is $2\Delta t$, and taking into account Pythagoras' theorem, we
have
\begin{equation}
c^2(\Delta t)^2=v^2(\Delta t)^2+d^2  \,.
\end{equation}
Using $\Delta t'=d/c$, one finally deduces the following
relationship
\begin{equation}
\Delta t=\gamma \Delta t'\,.
  \label{timedilation}
\end{equation}
As $\gamma >1$, then $\Delta t>\Delta t'$, so that time as
measured by the moving reference frame ${\cal O}'$ slows down
relatively to ${\cal O}$.

The above relationship may also be deduced directly from the
Lorentz transformations. Suppose that a clock sits at rest with
respect to the inertial reference frame ${\cal O}'$, in which two
successive clicks, represented by two events $A$ and $B$ are
separated by a time interval $\Delta t'$. To determine the time
interval $\Delta t$ as measured by ${\cal O}$, it is useful to
consider the inverse Lorentz transformation, given by
\begin{equation}
t=\gamma (t'+vx'/c^2)\,,
\end{equation}
which provides
\begin{equation}
t_B-t_A=\gamma \left[t'_B-t'_A+v(x'_B-x'_A)/c^2\right]\,,
\end{equation}
where $t_A$ and $t_B$ are the two clicks measured in ${\cal O}$.
As the events are stationary relative to ${\cal O}'$, we have
$x'_B=x'_A$, so that one finally ends up with Eq.
(\ref{timedilation}), taking into account $\Delta t=t_B-t_A$ and
$\Delta t'=t_B'-t_A'$. We note that the fact that a moving clock
slows down is completely reciprocal for any pair of inertial
observers, and this is essentially explained as both disagree
about simultaneity.

\bigskip

\begin{figure}[h]
\centering
\includegraphics[width=2.0in]{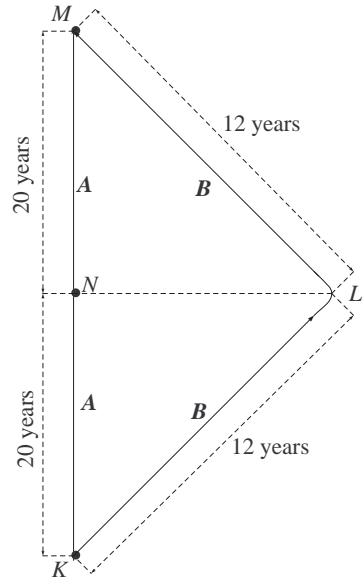}
\caption{Time dilation effects in the twin paradox. See the text
for details.} \label{twin}
\end{figure}

An interesting example of the time dilation effects is the
so-called ``twin paradox'', depicted in Fig. \ref{twin}. Consider
two identical twins, $A$ and $B$, respectively, where $A$ remains
at rest, while $B$ travels away from $A$ at a relativistic
velocity, close to the speed of light \cite{Ellis:2006sq}. As a
practical example, consider that $B$ initially recedes away from
$A$ at a speed of $v=4c/5$ for 12 years, as measured by $B$'s
clock, then returns at the same speed for 12 years. Thus, $B$
measures a total journey time of 24 years. One may ask what the
total travel time is, as measured by the twin $A$. To answer this,
consider that the event $K$, relatively to the twin $A$, is where
$B$ begins his outward journey; $L$ is the event when $B$ turns
around; and $M$ the event when $B$ arrives back at $A$. From Eq.
(\ref{gamma}), we verify that in both the outward and return
journey, we have $\gamma=[1-(4/5)^2]^{-1/2}=5/3$. Considering that
the event $N$, relative to $A$, is simultaneous to $L$, then one
verifies that $t_{KN}=\gamma t'_{KL}=5/3\times 12=20$ years, so
that the total travel time as measured by $A$ is 40 years.

However, time dilation effects are reciprocal between two inertial
frames, and one may wonder how it is possible to reconcile the
difference between both observers. It is important to emphasize
that the difference between both observers, $A$ and $B$, is that
twin $B$ is not an inertial observer, as his trajectory consists
of two inertial segments joined by a period of acceleration. It is
interesting to note that this feature has been observed
experimentally, in particular, in the Hafele-Keating experiment
\cite{Hafele-Keating}, which we refer to below.

\subsection{Time in general relativity}

The analysis outlined above has only taken into account flat
spacetimes, contrary to Einstein's general theory of relativity,
in which gravitational fields are represented through the
curvature of spacetime. In the discussion of special relativity,
the analysis was restricted to inertial motion, and in general
relativity the principle of relativity is extended to all
observers, inertial or non-inertial. In general relativity it is
assumed that the {\it the laws of physics are the same for all
observers, no matter what their state of motion} \cite{Ellis}. Now
it is clear that a gravitational force measured by an observer
essentially depends on his state of acceleration, which leads to
the {\it principle of equivalence}, which states that ``there is
no way of distinguishing between effects on an observer of a
uniform gravitational field and of constant acceleration.''

\begin{figure}[h]
\centering
\includegraphics[width=2.4in]{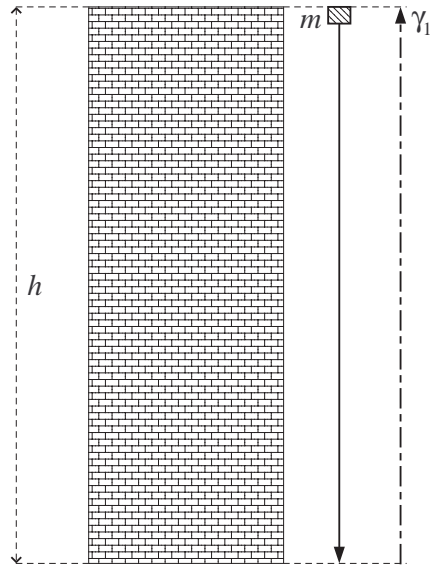}
\caption{Figure representing the gravitational time dilation
effect. Suppose that a particle of mass $m$ is dropped from the
top of a tower, falling freely with acceleration $g$ and reaching
the ground with velocity $v=(2gh)^{1/2}$. The idealized particle
is then converted into a single photon $\gamma_1$ and returns to
the top of the tower. An observer, at the top of the tower,
measures the time interval $\Delta t'=\Delta t (1+gh/c^2)$, where
$\Delta t$ is the time interval as measured by an observer at the
bottom of the tower. Thus, this shows that time flows at a faster
rate on top of the tower than at the bottom. See the text for
details.} \label{plot-redshift}
\end{figure}

The general theory of relativity has been an extremely successful
theory, with a well established experimental footing, at least for
weak gravitational fields. Of particular interest in this work are
the gravitational time dilation effects. For this, imagine the
following idealized thought experiment, suggested by Einstein
\cite{Schutz}, which is depicted in Fig. \ref{plot-redshift}.
Consider a tower of height $h$ hovering on the Earth's surface,
with a particle of rest mass $m$ lying on top. The particle is
then dropped from rest, falling freely with acceleration $g$ and
reaches the ground with a non-relativistic velocity
$v=(2gh)^{1/2}$. Thus, an observer on the ground measures its
energy as
\begin{equation}\label{energy}
E=mc^2+\frac{1}{2}mv^2=mc^2+mgh \,.
\end{equation}
The idealized particle is then converted into a single photon
$\gamma_1$ with identical energy $E$, which returns to the top of
the tower. Upon arrival it converts into a particle with energy
$E'=m'c^2$. Note that to avoid perpetual motion $m'>m$ is
forbidden, so that we consider $m=m'$, and the following
relationship is obtained
\begin{equation}\label{E-ratio}
\frac{E'}{E}=\frac{mc^2}{mc^2+mgh}\simeq 1-\frac{gh}{c^2} \,,
\end{equation}
with $gh/c^2\ll 1$. From the definitions $E=h\nu$ and $E'=h\nu'$,
where $\nu$ and $\nu'$ are the frequencies of the photon at the
bottom and top of the tower, so that from Eq. (\ref{E-ratio}), one
obtains
\begin{equation}\label{nu-ratio}
\nu'=\nu \left(1-\frac{gh}{c^2}\right) \,.
\end{equation}
This is depicted in Fig. \ref{plot-redshift}.
\bigskip

Now, to obtain the important result that clocks run at different
rates in a gravitational field, consider the following thought
experiment. The observer at the bottom of the tower emits a light
wave, directed to the top. The relationship of time between two
crests is simply the inverse of the frequency, i.e., $\Delta
t=1/\nu$, so that from Eq. (\ref{nu-ratio}), one obtains the
approximations, considering $gh/c^2 \ll 1$,
\begin{equation}\label{GRtimedilate}
\Delta t'=\Delta t \left(1+\frac{gh}{c^2}\right) \,.
\end{equation}
This provides the result that time flows at a faster rate on top
of the tower than at the bottom. Note that this result has been
obtained independently of the gravitational theory.

\subsection{Experimental tests}

A well-known experiment to test the time dilation effects in
general relativity, in particular, that clocks should run at
different rates at different places in a gravitational field, is
the Pound-Rebka experiment \cite{Pound-Rebka}, which confirmed the
predictions of general relativity to a $10\%$ precision level
\cite{Pound-Rebka2}. These results were later improved to a $1\%$
precision level by Pound and Snider \cite{Pound-Snider}.

The Hafele-Keating experiment \cite{Hafele-Keating}, realized in
October 1971 was an interesting test of the theory of relativity.
It essentially consisted of travelling four cesium-beam atomic
clocks aboard commercial airliners, and flying twice around the
world, first eastward, then westward. The results were then
compared with the clocks of the United States Naval Observatory.
To within experimental error, the results were consistent with the
relativistic predictions.

A modern application of the special and general relativistic time
dilation effects are the synchronization of atomic clocks on board
the Global Positioning System (GPS) satellites. The GPS has become
a widely used aid to navigation worldwide, enabling a GPS receiver
to determine its location, speed and direction. Now, as verified
above, general relativity predicts that the atomic clocks at GPS
orbital altitudes will tick more rapidly, as they are in a weaker
gravitational field than atomic clocks on Earth's surface; whilst
atomic clocks moving at GPS orbital speeds will tick more slowly
than stationary ground clocks, as predicted by special relativity.
When both effects are combined, the experimental data shows that
the on-board atomic clock rates do indeed agree with ground clock
rates to the predicted extent.

\section{Time irreversibility and the arrow of time}

The modern perspective in physics is that the Universe is
essentially governed by ``dynamical laws'', i.e., they specify the
evolution of a determined physical system with time, given the
initial conditions of the physical state. One normally considers
the evolution of physical systems into the future, which are
governed by differential equations \cite{Penrose}. In this
context, it is not common practice to evolve the physical systems
into the past, despite the fact that the dynamical equations of
classical and quantum physics are symmetrical under a time
reversal. Mathematically, one might as well specify the final
conditions and evolve the physical system back in time. One
specifies data at some initial instant and these data evolve,
through dynamical equations, to determine the physical state of
the system in the future, or to the past, i.e., detailed
predictability to the future and past is in principle possible.
However, several issues are raised by thermodynamics, general
relativity and quantum mechanics on time irreversibility and the
arrow of time.

\subsection{The arrow of time in thermodynamics}

A classical example of an irreversible process is the dissipation
of smoke from a lit cigarette. In principle, the evolution of the
system and its outcome is possible if the microscopic dynamics of
each individual particle is possible, but in practice one has
little knowledge of the position and velocity of every particle in
the system. The overall behavior  of the system is well described
in terms of appropriate averages of the physical parameters of the
individual particles, such as the distribution of mass, momentum
and of energy, etc. One may argue that the knowledge of these
averaged parameters is sufficient to determine the dynamical
behavior of the system and the respective final outcome. However,
this is not always the case, as in the specific examples of
`chaotic systems'.

Chaotic systems are classical systems, where a small change in the
initial conditions modifies the behavior of the system
exponentially, resulting in an unpredictability of the final
outcome. This chaotic unpredictability is closely related to the
Second Law of Thermodynamics, which states that the entropy of the
system increases (or at least does not decrease) with time. The
entropy is essentially a measure of the disorder or randomness in
the system. For instance, note the increased randomness of the
convoluted path of the diffusion of smoke from a lit cigarette. As
a simple example, consider a body moving through the air. The body
possesses kinetic energy, in an organized form, and as it slows
down from the air resistance, the kinetic energy has been
transferred to the random motion of the air particles and the
individual particles of the body \cite{Penrose}.

Consider the flow of heat from a hot body to a cooler body. The
evolution of this system is deterministic in character and
predicted by the Second Law of Thermodynamics. If one
theoretically considers the time-reversed evolution of the system,
then one would have the following scenario: Two bodies of the same
temperature evolve to bodies of unequal temperature. It would even
be a practical impossibility to know which body would be the
hotter, and which the cooler. Note that this difficulty of
dynamical retrodiction applies to most macroscopic systems, which
possess a large number of constituent particles and behaving in
accordance to the Second Law of Thermodynamics.

A particularly interesting example is that of friction
\cite{Ellis:2006sq}. Consider, for simplicity, a block of mass
sliding along a plane, and being slowed down by a constant force
of friction, consequently coming to rest at a determined instant.
One may agree that providing the detailed micro-physical
properties, such as the distribution of heat on the plane, it may
be possible to predict the initial conditions. However, using a
macroscopic viewpoint, after the system has settled down, one
cannot retrodict the initial conditions. One cannot even
reconstruct the trajectory, and one could conjecture if the block
came from the left or from the right.

In all of the examples outlined above one may argue that the
micro-physics is completely deterministic, contrary to the outcome
of macroscopic viewpoint \cite{Ellis:2006sq}. This is essentially
due to the fact that in the macroscopic viewpoint one does not
have enough detail of the physical system's micro-properties. One
may provide a statistical prediction of the eventual outcome, but
not a detailed and definite prediction. Note that the total energy
of the system is conserved as dictated by the First Law of
Thermodynamics. One may state that the disorder or randomness has
increased. Thus, the increase of entropy generically provides a
thermodynamic arrow of time. It is also possible to assume that
the Second Law of Thermodynamics and the thermodynamic arrow of
time are a consequence of the initial conditions of the universe,
which leads us to the cosmological arrow of time, that inexorably
points in the direction of the universe's expansion.

\subsection{The arrow of time in quantum mechanics}

Quantum uncertainty is a fundamental aspect of quantum theory,
i.e., it is not possible to determine a unique outcome of quantum
events. Formally, consider the wave function $\Psi(x)$ as a linear
combination of eigenfunctions $u_n(x)$, given by
\begin{equation}
\Psi_1(x)=\sum_i a_i u_i(x) \,.
  \label{initial}
\end{equation}
Suppose now that a measurement takes place at $t=\bar{t}$, thus
reducing the wave function to
\begin{equation}
\Psi_2(x)=a_n \, u_n(x) \,,
   \label{final}
\end{equation}
for some specific value $i=n$. It is important to emphasize that
the initial state (\ref{initial}) does not uniquely determine the
final state (\ref{final}). This is not due to lack of data, but is
due to the nature of quantum physics. Furthermore, one cannot
predict the final eigenstate $\Psi_2(x)$ from the initial state
\cite{Ellis:2006sq}. One cannot also retrodict to the past at the
quantum level, as once the wave function has collapsed to an
eigenstate, one cannot know the initial state from the final
state. Thus, despite the fact that there is time-symmetry in the
evolution of a quantum system, the reduction of the wave function
is essentially time-asymmetric.

\begin{figure}[h]
\centering
\includegraphics[width=3.0in]{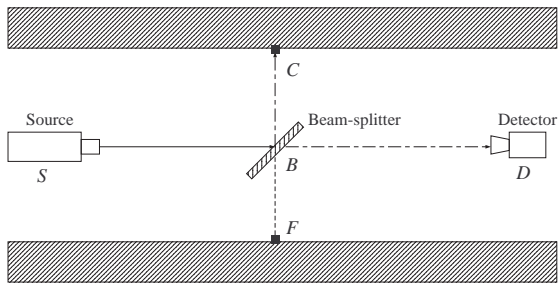}
\caption{Time asymmetry in the reduction of the wave function. See
the text for details.} \label{plot-reduction}
\end{figure}

It is illustrative to consider the following example
\cite{Penrose}. Consider a photon source $S$ which emits
individual photons. The latter are aimed at a beam-splitter $B$,
which is simply a half-silvered mirror, and is placed at an angle
of $45^{o}$ to the beam. Thus, if a photon is reflected, it will
be absorbed at the ceiling $C$; if it is transmitted, it will
activate a detector $D$. Suppose that the probability of
reflection and transmission is 50\%, respectively. Now, suppose
that a detection at $D$ is verified, which is equivalent to the
reduction of the wave function corresponding to Eq. (\ref{final}).
Given this, one may ask what the initial probabilities are. For
this, the relevant histories would be $SBD$ and $FBD$, where $F$
is a point on the floor. If a photon were emitted at $F$, it would
be reflected at $B$ and be detected at $D$. Now, applying the
quantum mechanical rules, one verifies that there is a 50\%
probability that the photon be detected at $D$, for the respective
emissions at $S$ and at $F$. This is absurd, as there is a 0\%
probability that a photon would be emitted from $F$. From this
simple example, one verifies time asymmetry related to the
reduction of the wave function in quantum mechanics. This is
depicted in Fig. \ref{plot-reduction}.

\section{Closed timelike curves and causality violation}

As time is incorporated into the proper structure of the fabric of
spacetime, it is interesting to note that general relativity is
contaminated with non-trivial geometries which generate {\it
closed timelike curves} \cite{Visser}. A closed timelike curve
(CTC) allows time travel, in the sense that an observer which
travels on a trajectory in spacetime along this curve, returns to
an event which coincides with the departure. The arrow of time
leads forward, as measured locally by the observer, but globally
he/she may return to an event in the past. This fact apparently
violates causality, opening Pandora's box and producing time
travel paradoxes \cite{Nahin}, throwing a veil over our
understanding of the fundamental nature of time. The notion of
causality is fundamental in the construction of physical theories,
therefore time travel and its associated paradoxes have to be
treated with great caution. The paradoxes fall into two broad
groups, namely the {\it consistency paradoxes} and the {\it causal
loops}.

The consistency paradoxes include the classical grandfather
paradox. Imagine travelling into the past and meeting one's
grandfather. Nurturing homicidal tendencies, the time traveller
murders his grandfather, impeding the birth of his father,
therefore making his own birth impossible. Another example is that
of autoinfanticide, where the time traveller returns to the past,
and kills himself as a baby. In fact, there are many versions of
the grandfather paradox, limited only by one's imagination. The
consistency paradoxes occur whenever possibilities of changing
events in the past arise.

The paradoxes associated with causal loops are related to
self-existing information or objects, trapped in spacetime.
Imagine a researcher travelling forward in time and reading the
details of the recently formulated, and anxiously anticipated,
consistent theory of quantum gravity. Returning to his time, he
explains the details to an ambitious younger colleague, who writes
it up and the article is eventually published in the journal,
where the first researcher read it after travelling into the
future. The article on the theory of quantum gravity exists in the
future because it was written in the past by the young researcher.
The latter wrote it up, after receiving the details from his
colleague, who in turn read the article in the future. Both parts
considered by themselves are consistent, and the paradox appears
when considered as a whole. One is liable to ask, what is the
origin of the information, as it appears out of nowhere. The
details for a complete and consistent theory of quantum gravity,
which paradoxically were never created, nevertheless exist in
spacetime. Note the absence of causality violations in these
paradoxes.

A great variety of solutions to the Einstein field equations
containing closed timelike curves exist, but two particularly
notorious features seem to stand out \cite{Lobo:2002rp}. Solutions
with a tipping over of the light cones due to a rotation about a
cylindrically symmetric axis; and solutions that violate the
energy conditions of general relativity, which are fundamental in
the singularity theorems and theorems of classical black hole
thermodynamics \cite{Visser,Hawking}.

\subsection{Stationary, axisymmetric solutions}

\begin{figure}[h]
\centering
\includegraphics[width=2.2in]{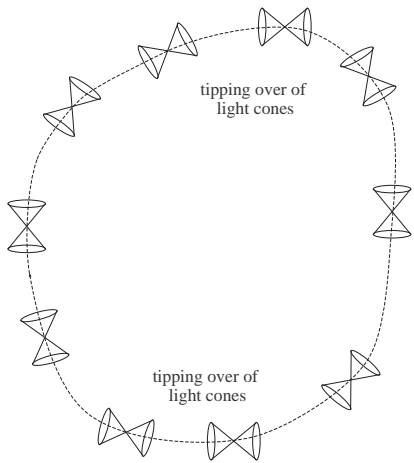}
\caption{The tipping over of light cones, depicted in the figure
is a generic feature of some solutions with a rotating cylindrical
symmetry. The dashed curve represents a closed timelike curve.}
\label{tipcones}
\end{figure}

The tipping over of light cones seems to be a generic feature of
some solutions with a rotating cylindrical symmetry, which is
depicted in Fig. \ref{tipcones}. The present work is far from
making an exhaustive search of all the Einstein field equation
solutions generating closed timelike curves with these features,
but the best known spacetimes will be briefly mentioned. The
earliest solution to the Einstein field equations containing
closed timelike curves, is probably that of the van Stockum
spacetime \cite{Visser,Tipler}. It is a stationary, cylindrically
symmetric solution describing a rapidly rotating infinite cylinder
of dust, surrounded by vacuum. The centrifugal forces of the dust
are balanced by the gravitational attraction. The light cones tip
over close to the cylinder, due to the strong curvature of the
spacetime, consequently inducing closed timelike curves. In 1949,
Kurt G\"{o}del discovered another exact solution to the Einstein
field equations consisting of a uniformly rotating universe
containing dust and a nonzero cosmological constant \cite{Godel}.
It is possible to show that moving away from the axis, the light
cones open out and tilt in the angular direction, eventually
generating closed timelike curves \cite{Hawking}.

An analogous solution to that of the van Stockum spacetime,
although possessing a different asymptotic behavior, is that of an
infinitely long straight string that lies and spins around the
$z$-axis~\cite{Visser}. These latter solutions also induce closed
timelike curves. An interesting variant of these rotating cosmic
strings is an extremely elegant model of a time-machine,
theoretically constructed by Gott \cite{Gott}. It is an exact
solution to the Einstein field equation for the general case of
two moving straight cosmic strings that do not intersect. This
solution produces closed timelike curves even though they do not
violate the weak energy condition, which essentially prohibits the
existence of negative energy densities, have no singularities and
event horizons, and are not topologically multiply-connected as
the wormhole solution, which will be considered below. The
appearance of closed timelike curves relies solely on the
gravitational lens effect and the relativity of simultaneity.
However, it was shown that the Gott time machine is unphysical in
nature, for such an acausal behaviour cannot be realized by
physical and timelike sources \cite{Deser,Deser2}.

\subsection{Solutions violating the energy conditions}

The traditional manner of solving the Einstein field equation
consists in considering a plausible distribution of energy and
matter, and then finding the geometrical structure. However, one
can run the Einstein field equation in the reverse direction by
imposing an exotic geometrical spacetime structure, and eventually
determine the matter source for the respective geometry.

In this fashion, solutions violating the energy conditions have
been obtained. One of the simplest energy conditions is the weak
energy condition, which is essentially equivalent to the
assumption that any timelike observer measures a local positive
energy density. Although classical forms of matter obey these
energy conditions, violations have been encountered in quantum
field theory, the Casimir effect being a well-known example.
Adopting the reverse philosophy, solutions such as traversable
wormholes \cite{Morris,Visser,phantomWH,Lemos:2003jb}, the warp
drive \cite{Alcubierre,Lobo:2004wq,Lobo:2002zf}, and the Krasnikov
tube \cite{Krasnikov} have been obtained. These solutions violate
the energy conditions and with simple manipulations generate
closed timelike curves \cite{MT,Everett,ER}.

We shall briefly consider the specific case of traversable
wormholes \cite{Morris}. A wormhole is essentially constituted by
two mouths, $A$ and $B$, residing in different regions of
spacetime~\cite{Morris}, which in turn are connected by a
hypothetical tunnel. One of the most fascinating aspects of
wormholes is their apparent ease in generating closed timelike
curves \cite{MT}. There are several ways to generate a time
machine using multiple wormholes \cite{Visser}, but a manipulation
of a single wormhole seems to be the simplest way \cite{MT}. The
basic idea is to create a time shift between both mouths. This is
done invoking the time dilation effects in special relativity or
in general relativity, i.e., one may consider the analogue of the
twin paradox, in which the mouths are moving one with respect to
the other, or simply the case in which one of the mouths is placed
in a strong gravitational field, so that time slows down in the
respective mouth~\cite{Visser,frolovnovikovTM}.

\begin{figure*}
\centering
  \includegraphics[width=2.3in]{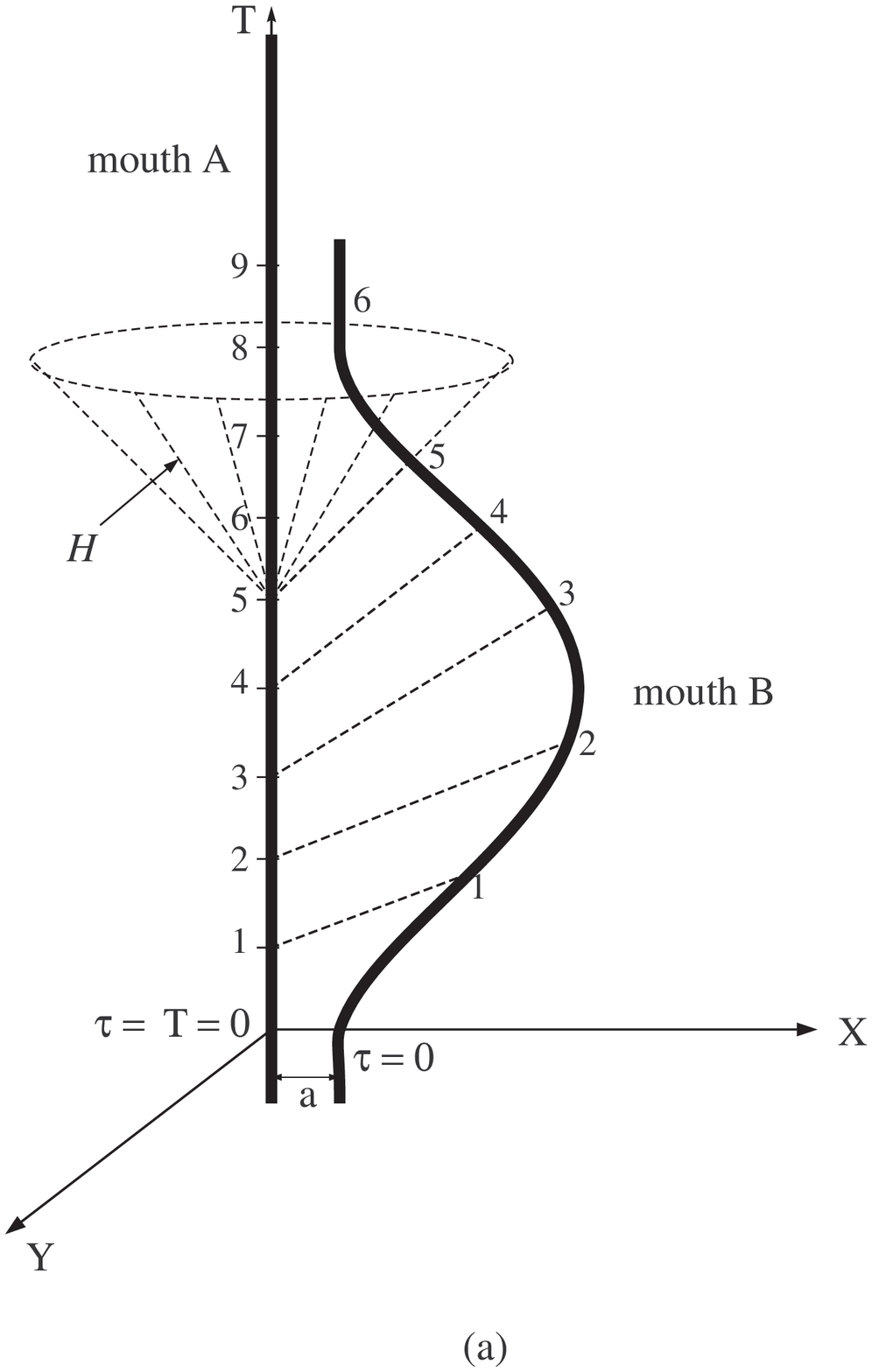}
  \hspace{0.4in}
  \includegraphics[width=2.3in]{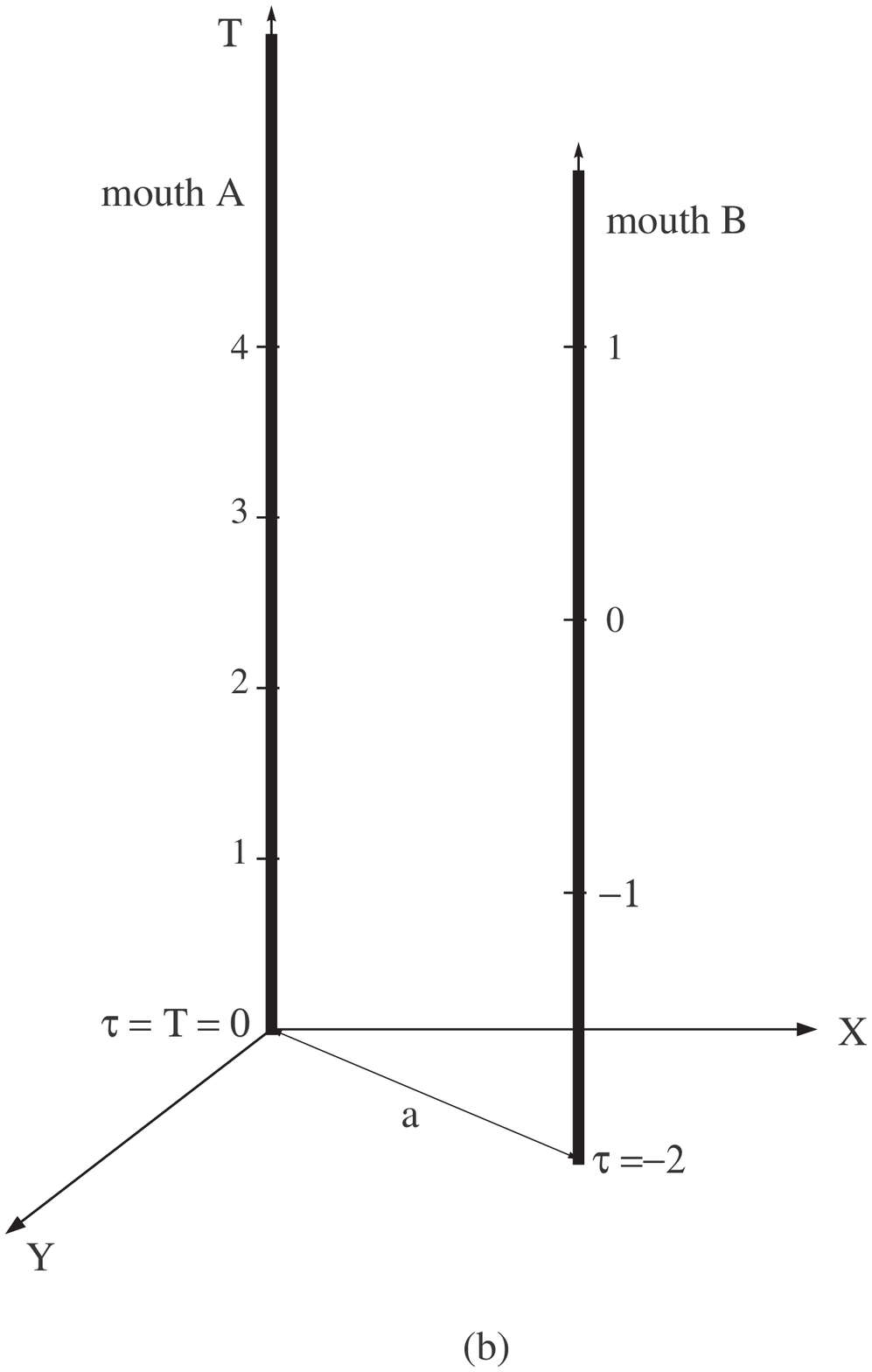}
  \caption[Wormhole spacetimes with closed timelike curves]
  {Depicted are two examples of wormhole spacetimes with closed
  timelike curves. The wormholes tunnels are arbitrarily short, and
  its two mouths move along two world tubes depicted as thick
  lines in the figure. Proper time $\tau$ at the wormhole throat is
  marked off, and note that identical values are the same event as seen
  through the wormhole handle. In Figure $(a)$, mouth $A$ remains at rest,
  while mouth $B$ accelerates from $A$ at a high velocity, then
  returns to its starting point at rest. A time shift is induced
  between both mouths, due to the time dilation effects of special
  relativity. The light cone-like hypersurface ${\it H}$ shown is
  a Cauchy horizon, beyond which predictability breaks down.
  Through every event to the future of ${\it H}$
  there exist closed timelike curves, and on the other hand there
  are no closed timelike curves to the past
  of ${\it H}$. In Figure $(b)$, a time shift between both mouths
  is induced by placing mouth $B$ in strong gravitational field.
  See text for details.}
  \label{fig:WH-time-machine}
\end{figure*}

To create a time shift using the twin paradox analogue, consider
that the mouths of the wormhole may be moving one with respect to
the other in external space, without significant changes of the
internal geometry of the tunnel. For simplicity, consider that one
of the mouths $A$ is at rest in an inertial frame, whilst the
other mouth $B$, initially at rest practically close by to $A$,
starts to move out with a high velocity, then returns to its
starting point. Due to the Lorentz time contraction, the time
interval between these two events, $\Delta T_B$, measured by a
clock comoving with $B$ can be made to be significantly shorter
than the time interval between the same two events, $\Delta T_A$,
as measured by a clock resting at $A$. Thus, the clock that has
moved has been slowed by $\Delta T_A-\Delta T_B$ relative to the
standard inertial clock. Suppose that the tunnel, between $A$ and
$B$ remains practically unchanged, so that an observer comparing
the time of the clocks through the handle will measure an
identical time, as the mouths are at rest with respect to one
another. However, by comparing the time of the clocks in external
space, he will verify that their time shift is precisely $\Delta
T_A-\Delta T_B$, as both mouths are in different reference frames,
frames that moved with high velocities with respect to one
another. Time is hooked up differently as measured through the
interior or in the exterior of the wormhole. Now, consider an
observer starting off from $A$ at an instant $T_0$, measured by
the clock stationed at $A$. He makes his way to $B$ in external
space and enters the tunnel from $B$. Consider, for simplicity,
that the trip through the wormhole tunnel is instantaneous. He
then exits from the wormhole mouth $A$ into external space at the
instant $T_0-(\Delta T_A-\Delta T_B)$ as measured by a clock
positioned at $A$. His arrival at $A$ precedes his departure, and
the wormhole has been converted into a time machine. See Figure
\ref{fig:WH-time-machine}.

\section{Summary and Discussion}

In this chapter, a brief review on the nature of time in Physics
has been explored (see Ref. \cite{Bertolami:2008ps} for a recent
review). In particular, it was noted that in Newtonian physics,
time flows at a constant rate for all observers, providing the
notion of absolute time, while in the special and general theories
of relativity, time necessarily flows at different rates for
different observers. It was shown that special relativity denies
the possibility of universal simultaneity, and consequently the
impossibility of a universal now. In this context, the Block
Universe description emerges, where all times, past and future are
equally present, and the notion of the flow of time is a
subjective illusion. This leads one to the possibility of time
being a dimension, contrary to a process. Indeed, intuitively, one
verifies that the notion of time as something that flows arises
due to an intimate relationship to change. Nevertheless, in
relativity the concept of spacetime being an independent entity
containing events predominates. Despite the fact of the great
popularity of the Block Universe representation in the Physics (in
particular the relativistic) community, this viewpoint often meets
with resistance, and we refer the reader to an interesting paper
by George F. Ellis \cite{Ellis:2006sq}.

Ellis argues that the Block Universe picture does not constitute a
realistic model of the Universe for several reasons: It assumes
simplified equations of state and thus does not apply to
spacetimes including complex systems, e.g., biological systems;
they don't take into account several issues such as dissipative
effects, feedback effects, and quantum uncertainty. Indeed, in our
everyday experience, psychological time does contrast to the Block
Universe picture, due to the subjective emergent notion of time as
a process. Thus, adopting this viewpoint, it is plausible to
consider that time is an abstract concept, non-existent as a
physical entity, but useful in describing processes. Contrary to
the Block Universe, representing spacetime as a fixed whole, Ellis
argues in favor of an Evolving Block Universe model of spacetime
\cite{Ellis:2006sq}, where ``... time progresses, events happen,
and history is shaped. Things could have been different, but
second by second, one specific evolutionary history out of all the
possibilities is chosen, takes place, and gets cast in stone''
\cite{Ellis:2006sq}. The Evolving Block Universe representation
defends that spacetime is extended into the future as events occur
along each worldline, which is determined by causal interactions.

A fundamental issue in the nature of time is its arrow. In modern
physics, dynamical laws essentially govern the Universe, where one
considers the evolution of physical systems into the future.
Nevertheless, the dynamical equations of classical and quantum
physics are symmetrical under a time reversal, and mathematically
one may evolve the physical systems into the past. In principle,
detailed predictability to the future and past is possible.
However, several issues are raised by thermodynamics and quantum
mechanics on time irreversibility and the arrow of time. In a
thermodynamical context, one may argue that the micro-physics of a
specific system is completely deterministic, contrary to the
outcome of the macroscopic viewpoint \cite{Ellis:2006sq}. This is
essentially due to the fact that in the macroscopic viewpoint one
does not have enough detail of the physical system's
micro-properties. One may provide a statistical prediction of the
eventual outcome, but not a detailed and definite prediction. This
fact is closely related to the Second Law of Thermodynamics. Thus,
the increase of entropy generically provides a thermodynamic arrow
of time. It is also possible to assume that the Second Law of
Thermodynamics and the thermodynamic arrow of time are a
consequence of the initial conditions of the universe, which leads
us to the cosmological arrow of time, that inexorably points in
the direction of the universe's expansion. In a quantum mechanical
context, it was also shown that despite the fact that there is a
time-symmetry in the evolution of a quantum system, the reduction
of the wave function is essentially time-asymmetric.

Relatively to causality violation, if one regards that general
relativity is a valid theory, then it is plausible to at least
include the {\it possibility} of time travel in the form of closed
timelike curves. However, a typical reaction is to exclude time
travel due to the associated paradoxes, although the latter do not
prove that time travel is mathematically or physically impossible.
The paradoxes do indeed indicate that local information in
spacetimes containing closed timelike curves is restricted in
unfamiliar ways. Relatively to the grandfather paradox, it is
logically inconsistent that the time traveller murders his
grandfather. But, one can ask, what exactly impeded him from
accomplishing his murderous act if he had ample opportunities and
the free-will to do so. It seems that certain conditions in local
events are to be fulfilled, for the solution to be globally
self-consistent. These conditions are denominated {\it consistency
constraints} \cite{Earman}. Much has been written on two possible
remedies to the paradoxes, namely the Principle of
Self-Consistency and the Chronology Protection Conjecture.

The Principle of Self-Consistency stipulates that events on a
closed timelike curve are self-consistent, i.e., events influence
one another along the curve in a cyclic and self-consistent way.
In the presence of closed timelike curves the distinction between
past and future events are ambiguous, and the definitions
considered in the causal structure of well-behaved spacetimes
break down. What is important to note is that events in the future
can influence, but cannot change, events in the past. According to
this principle, the only solutions of the laws of physics that are
allowed locally, and reinforced by the consistency constraints,
are those which are globally self-consistent.
Hawking's Chronology Protection Conjecture is a more conservative
way of dealing with the paradoxes. Hawking notes the strong
experimental evidence in favour of the conjecture from the fact
that ``we have not been invaded by hordes of tourists from the
future'' \cite{Hawk2}. An analysis reveals that the value of the
renormalized expectation quantum stress-energy tensor diverges
close to the formation of closed timelike curves, which destroys
the wormhole's internal structure before attaining the Planck
scale. There is no convincing demonstration of the Chronology
Protection Conjecture, but perhaps an eventual quantum gravity
theory will provide us with the answers.

But, as stated by Thorne \cite{Thorne}, it is by extending the
theory to its extreme predictions that one can get important
insights to its limitations, and probably ways to overcome them.
Therefore, time travel in the form of closed timelike curves, is
more than a justification for theoretical speculation, it is a
conceptual tool and an epistemological instrument to probe the
deepest levels of general relativity and extract clarifying views.
Relative to the issue of time, one may consider that the
underlying question is that of an ontological nature, and quoting
Ellis \cite{Ellis}: ``Does spacetime indeed exist as a real
physical entity, or is it just a convenient way of describing
relationships between physical objects, which in the end are all
that really exist at a fundamental level?''

\bigskip

\acknowledgments I thank Simon Grondin for the kind invitation to
write this chapter; Giuseppe de Risi and David Coule for extremely
stimulating discussions; and acknowledge funding from the
Funda\c{c}\~{a}o para a Ci\^{e}ncia e a Tecnologia (FCT)--Portugal
through the grant SFRH/BPD/26269/2006. This work is dedicated to
my parents, wherever or whenever they may be.


\end{document}